\documentclass[5p]{elsarticle}
\usepackage[utf8]{inputenc}
\usepackage{graphicx}
\usepackage{epstopdf}
\usepackage{algorithm}
\usepackage[]{algpseudocode}
\usepackage{multirow}
\usepackage{mathtools}
\usepackage{tabu}
\usepackage{setspace}
\usepackage[caption=false,font=footnotesize]{subfig}
\usepackage[hyphens]{url}
\usepackage{lineno}
\usepackage[ 
  bookmarks = true,
  bookmarksopen = true,
  bookmarksnumbered = true,
  breaklinks = true,
  linktocpage,
  pagebackref = false,
  colorlinks = true,
  linkcolor = blue,
  urlcolor  = blue,
  citecolor = red,
  anchorcolor = green,
  hyperindex = true,
  hyperfigures
  ]{hyperref}

\journal{Computer Communications}

\makeatletter
\newcommand{\StatexIndent}[1][3]{%
  \setlength\@tempdima{\algorithmicindent}%
  \Statex\hskip\dimexpr#1\@tempdima\relax}

\newcommand*{\algrule}[1][\algorithmicindent]{\makebox[#1][l]{\hspace*{.5em}\vrule height .75\baselineskip depth .25\baselineskip}}%

\newcount\ALG@printindent@tempcnta
\def\ALG@printindent{%
    \ifnum \theALG@nested>0
        \ifx\ALG@text\ALG@x@notext
            \addvspace{-3pt}
        \else
            \unskip
            \ALG@printindent@tempcnta=1
            \loop
                \algrule[\csname ALG@ind@\the\ALG@printindent@tempcnta\endcsname]%
                \advance \ALG@printindent@tempcnta 1
            \ifnum \ALG@printindent@tempcnta<\numexpr\theALG@nested+1\relax
            \repeat
        \fi
    \fi
    }%
\usepackage{etoolbox}
\patchcmd{\ALG@doentity}{\noindent\hskip\ALG@tlm}{\ALG@printindent}{}{\errmessage{failed to patch}}
\makeatother

\newcommand{\TheName}{$\mu$Nap}

\newcommand{\PSalgorithm}[1]{
\begin{algorithm}[#1]
 \setstretch{1}
 \caption{\TheName{} implementation: main loop modification for energy saving during packet overhearing.}
 \label{alg:unap}
 \begin{algorithmic}[1]
 \State ... \Comment {Initialisation}
 \State \textbf{global} $C \gets$ \textbf{true} \Comment {Contention flag}
 \Loop \Comment {Main loop}
  \State ...
  \While {bytes remaining} \Comment {Receiving loop}
    \State \Call {Read}{}
    \If {$R_A = $ BSSID OR ($T_A = $ BSSID AND
    \StatexIndent[2] \quad $R_A$ is other unicast MAC) }
      \State \Call {Set\_Sleep}{$\Delta t_\mathrm{DATA}, \Delta t_\mathrm{NAV}$}
    \EndIf
  \EndWhile
  \State \Call {Check\_FCS}{} \Comment {Frame received}
  \If {is Beacon AND $\Delta t_\mathrm{NAV}>0$} \Comment {CFP starts}
    \State $C \gets$ \textbf{false}
  \ElsIf {is CF\_End} \Comment {CFP ends}
    \State $C \gets$ \textbf{true}
  \EndIf
  \State ...
 \EndLoop
 \Procedure {Set\_Sleep}{$\Delta t_\mathrm{DATA}, \Delta t_\mathrm{NAV}$}
  \State $\Delta t_\mathrm{sleep} \gets \Delta t_\mathrm{DATA} + \Delta t_\mathrm{SIFS}$
  \If {$C$ AND is not CTS AND $\Delta t_\mathrm{NAV} \le 32 767$}
    \State $\Delta t_\mathrm{sleep} \gets \Delta t_\mathrm{sleep} + \Delta t_\mathrm{NAV}$
  \EndIf
  \If {$\Delta t_\mathrm{sleep} \ge \Delta t_\mathrm{sleep,min}$}
    \State \Call{Sleep}{$\Delta t_\mathrm{sleep}$}
    \State \Call{Wait}{$\Delta t_\mathrm{DIFS} - \Delta t_\mathrm{SIFS}$}
    \State \textbf{go to} Main loop
  \EndIf
  \State \textbf{go to} Receiving loop
 \EndProcedure
 \end{algorithmic}
\end{algorithm}
}

\begin{document}

\begin{frontmatter}

\title{\TheName{}: Practical micro-sleeps for 802.11 WLANs}

\author[uc3m,imdea]{Arturo~Azcorra}
\ead{azcorra@it.uc3m.es}

\author[uc3m]{Iñaki~Ucar\corref{corrauth}}
\cortext[corrauth]{Corresponding author}
\ead{inaki.ucar@uc3m.es}

\author[unibs]{Francesco~Gringoli}
\ead{francesco.gringoli@unibs.it}

\author[uc3m,imdea]{Albert~Banchs}
\ead{banchs@it.uc3m.es}

\author[uc3m]{Pablo~Serrano}
\ead{pablo@it.uc3m.es}

\address[uc3m]{Universidad Carlos III de Madrid, 28911 Leganés, Spain}
\address[imdea]{IMDEA Networks Institute, 28918 Leganés, Spain}
\address[unibs]{Università degli Studi di Brescia, 25123 Brescia, Italy}

\tnotetext[copy]{\textcopyright 2017. This manuscript version is made available under the \href{http://creativecommons.org/licenses/by-nc-nd/4.0/}{CC-BY-NC-ND 4.0 license}. DOI: \href{http://doi.org/10.1016/j.comcom.2017.06.008}{10.1016/j.comcom.2017.06.008}}

\begin{abstract}

In this paper, we revisit the idea of putting interfaces to sleep during \emph{packet overhearing} (i.e., when there are ongoing transmissions addressed to other stations) from a practical standpoint. To this aim, we perform a robust experimental characterisation of the timing and consumption behaviour of a commercial 802.11 card. We design \TheName{}, a local standard-compliant energy-saving mechanism that leverages micro-sleep opportunities inherent to the CSMA operation of 802.11 WLANs. This mechanism is backwards compatible and incrementally deployable, and takes into account the timing limitations of existing hardware, as well as practical CSMA-related issues (e.g., capture effect). According to the performance assessment carried out through trace-based simulation, the use of our scheme would result in a 57\% reduction in the time spent in overhearing, thus leading to an energy saving of 15.8\% of the activity time.

\end{abstract}


\end{frontmatter}

\section{Introduction}\label{sec:intro}

IEEE 802.11 is the standard \emph{de facto} for broadband Internet access. The recent 802.11ac amendment opens up new opportunities by bringing Gigabit to wireless local area networks (WLANs). Since the seminal work \cite{Feeney2001}, energy efficiency stands as a major issue due to the intrinsic CSMA mechanism, which forces the network card to stay active performing \emph{idle listening}.

The 802.11 standard developers are fully aware of the energy issues that WiFi poses on battery-powered devices and have designed mechanisms to reduce energy consumption. One of such mechanisms is the Power Save (PS) mode, which is widely deployed among commercial wireless cards, although unevenly supported in software drivers. With this mechanism, a station (STA) may enter a doze state during long periods of time, subject to prior notification, if it has nothing to transmit. Meanwhile, packets addressed to this dozing STA are buffered and signalled in the Traffic Indication Map (TIM) attached to each beacon frame.

The PS mechanism dramatically reduces the power consumption of a wireless card. However, the counterpart is that, since the card is put to sleep for hundreds of milliseconds, the user experiences a serious performance degradation because of the delays incurred. The automatic power save delivery (APSD) introduced by the 802.11e amendment (\cite{Perez-Costa2010} gives a nice overview) is based on this mechanism, and aims to improve the downlink delivery by taking advantage of QoS mechanisms, but has not been widely adopted.

More recently, the 802.11ac amendment improves the PS capabilities with the VHT TXOP power save mechanism. Basically, an 11ac STA can doze during a TXOP (transmission opportunity) in which it is not involved. This capability is announced within the new VHT (Very High Throughput) framing format, so that the AP knows that it cannot send traffic to those STAs until the TXOP's natural end, even if it is interrupted earlier. Still, the potential dozing is in the range of milliseconds and may lead to channel underuse if these TXOPs are not fully exploited.

Considering shorter timescales, packet overhearing (i.e., listening to the wireless while there is an ongoing transmission addressed to other station) has been identified as a potential source of energy wastage \cite{basu2004}. Despite this, we have performed an extensive measurement campaign and have not found any attempt from manufacturers\footnote{Using our setup described in Section~\ref{sec:transition-times}, we have tested cards from different manufacturers with the latest available firmwares and drivers: Broadcom BCM43224, Realtek RTL8191SEvB, Atheros AR9280, Intel Wireless-AC 7260 and Qualcomm Atheros QCA9880, which is a very recent state-of-the-art 11ac card.} to implement solutions to lessen its impact.

In this work, we revisit this idea of packet overhearing as a trigger for sleep opportunities, and we take it one step further to the range of microseconds. To this end, we experimentally explore the timing limitations of 802.11 cards and, building on this knowledge, we design \TheName{}, a local standard-compliant energy-saving mechanism for 802.11 WLANs. With \TheName{}, a STA is capable of saving energy during packet overhearing autonomously, with full independence from the 802.11 capabilities supported or other power saving mechanisms in use, which makes it backwards compatible and incrementally deployable.

In summary, the main contributions of this paper are the following:

\begin{itemize}
 \item A robust experimental characterisation of the timing and consumption behaviour of a COTS (commercial off-the-shelf) wireless card.
 \item Design of \TheName{}, a local standard-compliant energy-saving mechanism that takes into account these timing limitations, as well as practical CSMA-related issues (e.g., capture effect, hidden nodes) not considered in prior work.
 \item Performance evaluation of \TheName{} based on our measurements and real wireless traces, and performance degradation analysis due to channel errors.
 \item Discussion of the impact and applicability of our mechanism. We draw attention to the need for standardising the hardware capabilities in terms of energy in 802.11.
\end{itemize}

The remainder of this paper is organised as follows. Section~\ref{sec:related} reviews related work. Section~\ref{sec:transition-times} experimentally explores the timing limitations of 802.11 COTS devices. Section~\ref{sec:unap} analyses micro-sleep opportunities in CSMA as well as several practical issues of 802.11 networks, and proposes \TheName{}. Section~\ref{sec:results} presents the performance evaluation and Section~\ref{sec:conclusions} summarises the conclusions of this paper.

\section{Related work}\label{sec:related}

It has been empirically proved that, in any network, the so-called power-law distribution, also known as Pareto distribution, holds for the traffic generated by its nodes. In other words, typically a few heavy hitters-generate most of the traffic while the majority of the nodes are only responsible for a small fraction, and this is true regardless of the network load. This means that the majority of nodes within any WLAN would spend most of the time in idle state.

There are two main strategies to save energy in this idle time: the first one targets \emph{idle listening} (the wireless channel is empty), and the second one targets \emph{packet overhearing} (there are other nodes communicating). To support these savings, COTS devices have two main operational states as a function of the reference clock used: the active state and the sleep state. The more a card stays in sleep state, the less power it consumes.

Since its conception, 802.11 has attempted to minimise idle listening with the introduction of the PS mode, and some previous work followed this path. For instance, Liu and Zhong \cite{Liu2008} proposed $\mu$PM to exploit short idle intervals ($<$100 ms) without buffering or cooperation. $\mu$PM predicts the arrival time of the next frame and puts the interface in PS mode while no arrivals are expected. This mechanism demonstrated poor granularity (tens of ms) on existing hardware and leads to performance degradation due to frame loss. Therefore, it is only suitable for low-traffic scenarios.

Others propose a PS-like operation. Jang \textit{et al.} \cite{Jang2011} described Snooze, an access point (AP)-directed micro-sleep scheduling and antenna configuration management method for 11n WLANs. As a consequence of its centralised design, the granularity of the so-called micro-sleeps in this approach is poor (few milliseconds), which poses doubts on its performance under heavy loads.

Zhang and Shin \cite{Zhang2012} addressed the issue from a different standpoint with their Energy-Minimizing Idle Listening (E-MiLi). E-MiLi adaptively downclocks the card during idle periods, and reverts to full rate when an incoming frame is detected. To achieve this purpose, they need to change the physical layer (PHY) all the way down to enable downclocked detection, which severely limits the potential gains. For instance, the E-MiLi downclocking factor of 16 would yield a high power consumption in a modern card compared to its sleep state (see \ref{sec:characterisation2}).

On the other hand, all indicators show that we should expect an exponential grow in the number of wireless devices connected. Thus, there is a rough consensus about that \emph{densification} will become one of the main aspects of next-generation wireless networks, which brings us back to the problem of packet overhearing. In this way, the recent 11ac amendment adds the ability to save energy during TXOPs, but this mechanism is restricted to QoS traffic, and the potential sleeps are coarse, in the range of milliseconds. Any sub-millisecond approach must take into account the timing parameters of the hardware. In fact, some early studies realise the importance of this issue when WiFi technology began to take off commercially \cite{kamerman1997, havinga2000, jung2002}.

Baiamonte and Chiasserini \cite{Baiamonte2006} were the first to chase fine-grained micro-sleep opportunities during packet overhearing. They define the Energy-efficient Distributed Access (EDA) scheme, which uses the 802.11 virtual carrier-sensing mechanism for power-saving purposes. Basically, a STA dozes when the Network Allocation Vector (NAV) or the backoff counter are non-zero. Unfortunately, this work lacks an empirical characterisation of the timing constraints needed to design a practical mechanism. Moreover, dozing during the backoff window is not 802.11-fair: in 802.11, STAs must sense the channel every single time slot during the contention period and, if another STA seizes the channel first, the backoff timer must be stopped in order to receive the incoming frame and set the NAV to the proper value. The EDA scheme allows STAs to doze during the contention period and, therefore, breaks the CSMA operation.

Balaji \textit{et al.} \cite{Balaji2010} revisited the problem of packet overhearing with a scheme called Sleep during Neighbor-Addressed Frame (SNAF). With SNAF, a wireless card checks the destination MAC address and switches to sleep state during the payload duration if it was addressed to other host. They assume, without any experimental validation though, an instantaneous switch-off and that the time required to wake up is equivalent to a Short Interframe Space (SIFS). In order to prevent the risk resulting from errors in the frame header that would lead to an incorrect NAV counter, the authors propose to introduce a new framing format with a new FCS devoted to the MAC header only. This solution lacks compatibility and introduces more overhead based on no evidence.

Building on the same idea, Sudarshan \textit{et al.} \cite{Prasad2014} proposed Übersleep. This time, the authors do not consider it necessary to add any extra FCS, as they claim (without any specific basis) that such errors are very unlikely.

More recently, Palacios-Trujillo \textit{et al.} modified DCF \cite{palacios2013a} and PCF \cite{palacios2013b} to exploit per-packet sleeps. They also applied these ideas to network coding \cite{palacios2015b} and to a polling-based version of 11ac's TXOP PS mode \cite{palacios2015a}. Unfortunately, all these papers rely on these early studies mentioned before \cite{kamerman1997, havinga2000, jung2002}, which analysed old wireless cards unable to perform sub-millisecond transitions between states.

\section{State transition times on 802.11 cards}\label{sec:transition-times}

\begin{figure}[t]
    \centering
    \includegraphics[width=\linewidth]{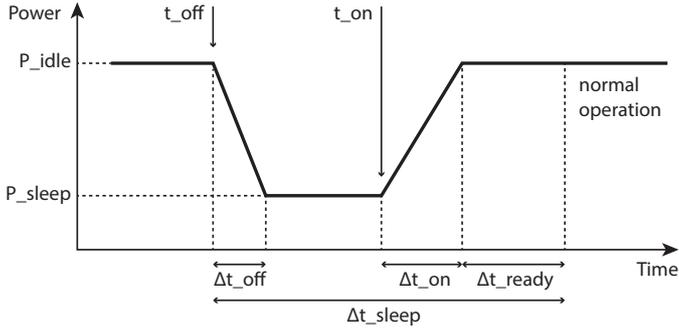}
    \caption{Generic sleep breakdown.}
    \label{fig:timing}
\end{figure}

From the hardware point of view, the standard PS mechanism requires supporting two states of operation: the \emph{awake state} and the \emph{sleep state}. The latter is implemented using a secondary low-frequency clock. Indeed, it is well-known that the power consumption of digital devices is proportional to the clock rate \cite{Zhang2012}. In fact, other types of devices, such as microcontroller-based devices or modern general-purpose CPUs, implement sleep states in the same way.

For any microcontroller-based device with at least an idle state and a sleep state, one would expect the following behaviour for an ideal sleep. The device was in idle state, consuming $P_\mathrm{idle}$, when, at an instant $t_\mathrm{off}$, the sleep state is triggered and the consumption falls to $P_\mathrm{sleep}$. A secondary low-power clock decrements a timer of duration $\Delta t_\mathrm{sleep} = t_\mathrm{on} - t_\mathrm{off}$, and then the expiration of this timer triggers the wake-up at $t_\mathrm{on}$. The switching between states would be instantaneous and the energy saving would be
\begin{align}\label{ec:idealsleep}
 E_\mathrm{save} = (P_\mathrm{idle} - P_\mathrm{sleep}) \cdot \Delta t_\mathrm{sleep}
\end{align}

This estimate could be considered valid for a time scale in the range of tens of milliseconds at least, but this is no longer true for micro-sleeps. Instead, Fig.~\ref{fig:timing} presents a conceptual breakdown of a generic micro-sleep. After the sleep state is triggered at $t_\mathrm{off}$, it takes $\Delta t_\mathrm{off}$ before the power consumption actually reaches $P_\mathrm{sleep}$. Similarly, after the wake-up is triggered at $t_\mathrm{on}$, it takes some time, $\Delta t_\mathrm{on}$, to reach $P_\mathrm{idle}$. Finally, the circuitry might need some additional time $\Delta t_\mathrm{ready}$ to stabilise and operate normally. Thus, the most general expression for the energy saved in a micro-sleep is the following:
\begin{align}\label{ec:realsleep}
 E'_\mathrm{save} & = E_\mathrm{save} - E_\mathrm{waste} \\
 &=  (P_\mathrm{idle} - P_\mathrm{sleep}) \cdot (\Delta t_\mathrm{sleep} -\Delta t_\mathrm{ready}) \nonumber\\
 & - \int_{\Delta t_\mathrm{off} \cup \Delta t_\mathrm{on}} (P - P_\mathrm{sleep}) \cdot dt  \nonumber
\end{align}

\noindent where we have considered a general waveform $P(t)$ for the transients $\Delta t_\mathrm{off}$ and $\Delta t_\mathrm{on}$. $E_\mathrm{waste}$ represents an energy toll per sleep when compared to the ideal case.

\begin{figure}[t]
    \centering
    \includegraphics[width=0.8\linewidth]{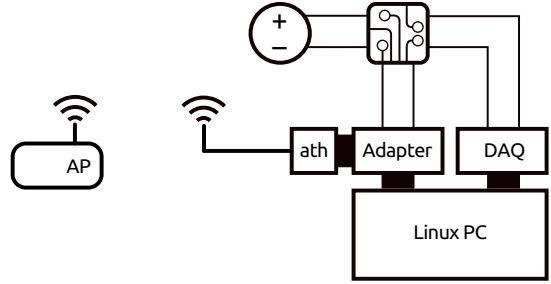}
    \caption{Measurement setup for energy performance characterisation.}
    \label{fig:setup}
\end{figure}

Our next objective is to quantify these limiting parameters, which can be defined as follows:

\begin{description}
\item[$\Delta t_\mathrm{off}$] is the time required to switch from idle power and to sleep power consumption.
\item[$\Delta t_\mathrm{on}$] is the time required to switch from sleep power to idle power consumption.
\item[$\Delta t_\mathrm{ready}$] is the time required for the electronics to stabilise and become ready to transmit/receive.
\end{description}

The sum of this set of parameters defines the minimum sleep time, $\Delta t_\mathrm{sleep,min}$, for a given device:
\begin{align}
 \Delta t_\mathrm{sleep,min} = \Delta t_\mathrm{off} + \Delta t_\mathrm{on} + \Delta t_\mathrm{ready}
\end{align}

Performing this experimental characterisation requires the ability to timely trigger the sleep mode on demand. Most COTS cards are not suitable for this task, because they implement all the low-level operations in an internal proprietary binary firmware. After an extensive study, we found that cards based on the open-source driver \texttt{ath9k} are well suited for our needs, as they do not load a firmware to operate, and the driver has access to very low level functionality (e.g., supporting triggering the sleep mode by just writing into a register). Leveraging on these properties, we conducted our experimental characterisation using an Atheros AR9280 Half-height Mini PCI Express card.

This card is attached to a PC through a flexible x1 PCI Express to Mini PCI Express adapter from Amfeltec, as the right part of Fig.~\ref{fig:setup} depicts. This adapter connects the PCI bus' data channels to the host and provides an ATX port so that the wireless card can be supplied by an external power source. The same PC holds a NI PCI-6289, a high-accuracy multifunction data acquisition (DAQ) device, optimised for 18-bit input accuracy. Its timing resolution is 50 ns with an accuracy of 50 ppm of sample rate. In this way, the operations sent to the wireless card and the energy measurements can be correlated using the same timebase. A small command-line tool was developed\footnote{\url{https://github.com/Enchufa2/daq-acquire}} to perform measurements on the DAQ card using the open-source Comedi\footnote{\url{http://comedi.org/}} drivers and libraries.

The power supply is a Keithley 2304A DC Power Supply, which is optimised for testing battery-operated wireless communication devices. It powers the wireless card through a measurement circuit that extracts the voltage and converts the current with a high-precision sensing resistor and amplifier. Considering that the DAQ card has a certain settling time, it can be modelled as a small capacitor which acts as a low-pass filter. Thus, two buffers (voltage followers) are placed before the DAQ card to decrease the output impedance of the measurement circuit \cite{ni2014}.

\begin{figure*}[t]
    \centering
    \subfloat[]{\label{fig:sleep-tx-a} \includegraphics[width=0.49\linewidth]{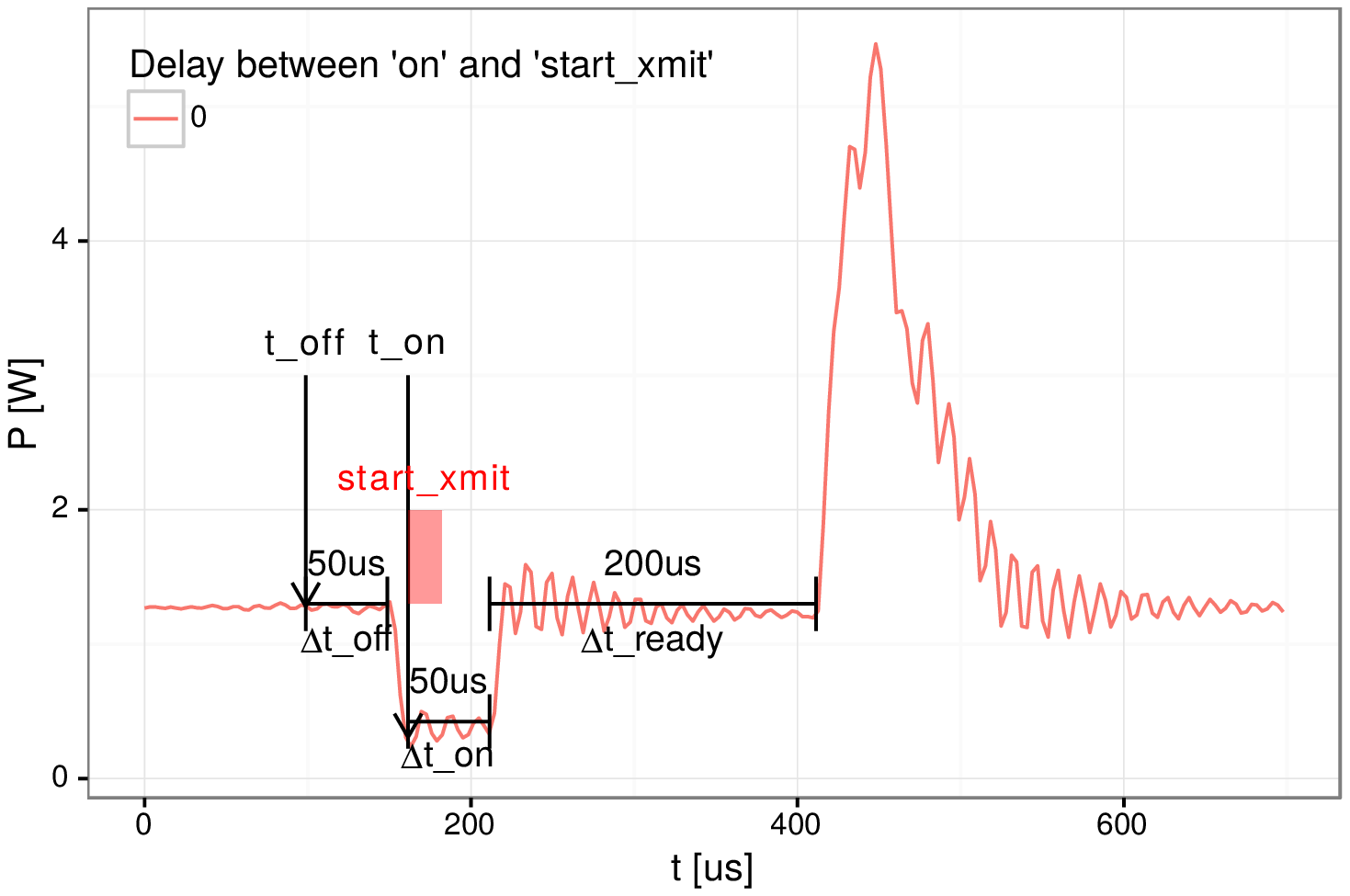}}
    \subfloat[]{\label{fig:sleep-tx-b} \includegraphics[width=0.49\linewidth]{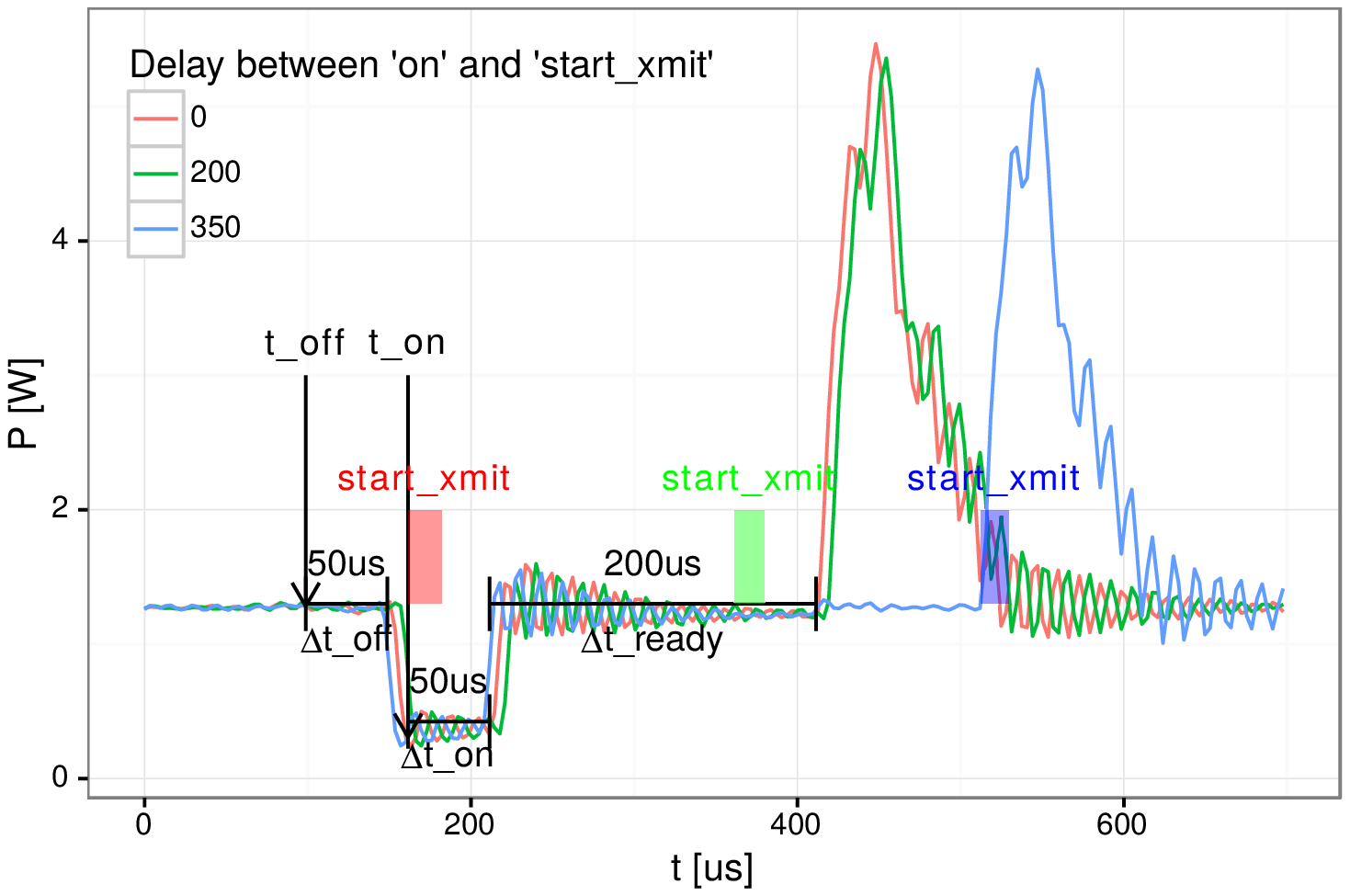}}
    \caption{Atheros AR9280 timing characterisation.}
\end{figure*}

The card under test is associated to an AP in 11a mode to avoid any interfering traffic from neighbouring networks. This AP is placed very close to the node to obtain the best possible signal quality, as we are simply interested in not losing the connectivity for this experiment. With this setup, the idea is to trigger the sleep state, then bring the interface back to idle and finally trigger the transmission of a buffered packet as fast as possible, in order to find the timing constraints imposed by the hardware in the power signature. From an initial stable power level, with the interface associated and in idle mode, we would expect a falling edge to a lower power level corresponding to the sleep state. Then the power level would raise again to the idle level and, finally, a big power peak would mark the transmission of the packet. By correlating the timestamps of our commands and the timestamps of the measured power signature, we are able to measure the limiting parameters $\Delta t_\mathrm{off}, \Delta t_\mathrm{on}, \Delta t_\mathrm{ready}$. 

The methodology to reproduce these steps required hacking the \texttt{ath9k} driver to timely trigger write operations in the proper card registers, and to induce a transmission of a pre-buffered packet directly in the device without going through the entire network stack. The code for reproducing this experiment is available on GitHub\footnote{\url{https://github.com/Enchufa2/crap/tree/master/ath9k/downup}}, and comprises the following steps:

\begin{enumerate}
\setcounter{enumi}{-1}
 \item Initially, the card is in idle state, connected to the AP.
 \item A RAW socket (Linux AF\_PACKET socket) is created and a socket buffer is prepared with a fake packet.
 \item $t_\mathrm{off}$ is triggered by writing a register in the card, which has proved to be almost instantaneous in kernel space.
 \item A micro-delay of 60 $\mu$s is introduced in order to give the card time to react.
 \item $t_\mathrm{on}$ is triggered with another register write.
 \item Another timer sets a programmable delay.
 \item The fake frame is sent using a low-level interface, i.e., calling the function \texttt{ndo\_start\_xmit()} from the \texttt{net\_device} operations directly. By doing this, we try to spend very little time in kernel.
\end{enumerate}

The power signature recorded as a result of this experiment is shown in Fig.~\ref{fig:sleep-tx-a}. As we can see, the card spends $\Delta t_\mathrm{off} = 50$ $\mu$s consuming $P_\mathrm{idle}$ and then it switches off to $P_\mathrm{sleep}$ in only 10 $\mu$s. Then, $t_\mathrm{on}$ is triggered. Similarly, the card spends $\Delta t_\mathrm{on} = 50$ $\mu$s consuming $P_\mathrm{sleep}$ and it wakes up almost instantaneously. Note that the transmission of the packet is triggered right after the $t_\mathrm{on}$ event and the frame spends very little time at the kernel (the time spent in kernel corresponds to the width of the rectangle labelled as \texttt{start\_xmit} in the graph). Nonetheless, the card sends the packet 200 $\mu$s after returning to idle, even though the frame was ready for transmission much earlier.

To understand the reasons for the delay in the frame transmission observed above, we performed an experiment in which frame transmissions were triggered at different points in time by introducing different delays between the $t_\mathrm{on}$ and \texttt{start\_xmit} events. Fig.~\ref{fig:sleep-tx-b} shows that the card starts transmitting always in the same instant whenever the kernel triggers the transmission within the first 250 $\mu$s right after the $t_\mathrm{on}$ event (lines 0 and 200). Otherwise, the card starts transmitting almost instantaneously (line 350). This experiments demonstrate that the device needs $\Delta t_\mathrm{ready} = 200$ $\mu$s to get ready to transmit/receive after returning to idle.

To sum up, our experiments show that, if we want to bring this card to sleep during a certain time $\Delta t_\mathrm{sleep}$, we should take into account that it requires a minimum sleep time $\Delta t_\mathrm{sleep,min}=300$ $\mu$s. Therefore, $\Delta t_\mathrm{sleep} \geq \Delta t_\mathrm{sleep,min}$ must be satisfied, and we must program the $t_\mathrm{on}$ interrupt to be triggered $\Delta t_\mathrm{on} + \Delta t_\mathrm{ready}=250$ $\mu$s before the end of the sleep. Note also that the card wastes a fixed time $\Delta t_\mathrm{waste}$ consuming $P_\mathrm{idle}$:
\begin{align}\label{ec:twaste}
 \Delta t_\mathrm{waste} = \Delta t_\mathrm{off} + \Delta t_\mathrm{ready}
\end{align}

\noindent which is equal to 250 $\mu$s also. Thus, the total time in sleep state is $\Delta t_\mathrm{sleep} - \Delta t_\mathrm{waste}$, and the energy toll from Equation~(\ref{ec:realsleep}) can be simplified as follows:
\begin{align}\label{ec:Ewaste}
 E_\mathrm{waste} \approx (P_\mathrm{idle} - P_\mathrm{sleep})\cdot\Delta t_\mathrm{waste}
\end{align}

\section{\TheName{} design}\label{sec:unap}

The key idea of \TheName{} is to put the interface to sleep during packet overhearing while meeting the constraint $\Delta t_\mathrm{sleep,min}$ identified in the previous section. Additionally, the algorithm should be local in order to be incrementally deployable, standard-compliant, and should take into account real-world practical issues. For this purpose, Section~\ref{sec:microsleep} analyses available micro-sleep opportunities in 802.11 and determines under which circumstances the NAV mechanism can be used to extend a micro-sleep while ensuring that no frames are lost within such time. Section~\ref{sec:issues} explores well-known practical issues of WLAN networks that had not been addressed by previous energy-saving schemes. Finally, Section~\ref{sec:design} presents \TheName{}.

\subsection{Micro-sleep opportunities in 802.11}\label{sec:microsleep}

Due to the CSMA mechanism, 802.11 STAs receive every single frame from their SSID or from others in the same channel (even some frames from overlapping channels). Upon receiving a frame, a STA checks the Frame Check Sequence (FCS) for errors and then, and only after having received the entire frame, it discards the frame if it is not the recipient. In 802.11 terminology, this is called \emph{packet overhearing}. Since packet overhearing consumes the power corresponding to a full packet reception that is not intended for the station, it represents a source of inefficiency. Thus, we could avoid this unnecessary power consumption by triggering micro-sleeps that bring the wireless card to a low-energy state.

Indeed, the Physical Layer Convergence Procedure (PLCP) carries the necessary information (rate and length) to know the duration of the PLCP Service Data Unit (PSDU), which consists of a MAC frame or an aggregate of frames. And the first 10 bytes of a MAC frame indicate the intended receiver, so a frame could be discarded very early, and the station could be brought to sleep if the hardware allows for such a short sleeping time. Therefore, the most naive micro-sleep mechanism could determine, given the constraint $\Delta t_\mathrm{sleep,min}$, whether the interface could be switched off in a frame-by-frame basis. And additionally, this behaviour can be further improved by leveraging the 802.11 virtual carrier-sensing mechanism. 

Virtual carrier-sensing allows STAs not only to seize the channel for a single transmission, but also to signal a longer exchange with another STA. For instance, this exchange can include the acknowledgement sent by the receiver, or multiple frames from a station in a single transmission opportunity (TXOP). So MAC frames carry a duration value that updates the Network Allocation Vector (NAV), which is a counter indicating how much time the channel will be busy due to the exchange of frames triggered by the current frame. And this duration field is, for our benefit, enclosed in the first 10 bytes of the MAC header too. Therefore, the NAV could be exploited to obtain substantial gains in terms of energy. 

In order to unveil potential sleeping opportunities within the different states of operation in 802.11, first of all we review the setting of the NAV. 802.11 comprises two families of channel access methods. Within the legacy methods, the Distributed Coordination Function (DCF) is the basic mechanism with which all STAs contend employing CMSA/CA with binary exponential backoff. In this scheme, the duration value provides single protection: the setting of the NAV value is such that protects up to the end of one frame (data, management) plus any additional overhead (control frames). For instance, this could be the ACK following a data frame or the CTS + data + ACK following an RTS.

When the Point Coordination Function (PCF) is used, time between beacons is rigidly divided into contention and contention-free periods (CP and CFP, respectively). The AP starts the CFP by setting the duration value in the beacon to its maximum value (which is 32 768; Table~8-3 of the IEEE Std 802.11-2012 \cite{80211} depicts the duration/ID field encoding). Then, it coordinates the communication by sending CF-Poll frames to each STA. As a consequence, a STA cannot use the NAV to sleep during the CFP, because it must remain CF-pollable, but it still can doze during each individual packet transmission. In the CP, DCF is used.

802.11e introduces traffic categories (TC), the concept of TXOP, and a new family of access methods called Hybrid Coordination Function (HCF), which includes the Enhanced Distributed Channel Access (EDCA) and the HCF Controlled Channel Access (HCCA). These two methods are the QoS-aware versions of DCF and PCF respectively.

Under EDCA, there are two classes of duration values: single protection, as in DCF, and multiple protection, where the NAV protects up to the end of a sequence of frames within the same TXOP. By setting the appropriate TC, any STA may start a TXOP, which is zero for background and best-effort traffic, and of several milliseconds for video and audio traffic as defined in the standard (Table~8-105 of the IEEE Std 802.11-2012 \cite{80211}). A non-zero TXOP may be used for dozing, as 11ac does, but these are long sleeps and the AP needs to support this feature, because a TXOP may be truncated at any moment with a CF-End frame, and it must keep buffering any frame directed to any 11ac dozing STA until the NAV set at the start of the TXOP has expired.

HCCA works similarly to PCF, but under HCCA, the CFP can be started at almost any time. In the CFP, when the AP sends a CF-poll to a STA, it sets the NAV of other STAs for an amount equal to the TXOP. Nevertheless, the AP may reclaim the TXOP if it ends too early (e.g., the STA has nothing to transmit) by resetting the NAV of other STAs with another CF-Poll. Again, the NAV cannot be locally exploited to perform energy saving during a CFP.

Finally, there is another special case in which the NAV cannot be exploited either. 802.11g was designed to bring the advantages of 11a to the 2.4 GHz band. In order to interoperate with older 11b deployments, it introduces CTS-to-self frames (also used by more recent amendments such as 11n and 11ac). These are standard CTS frames, transmitted at a legacy rate and not preceded by an RTS, that are sent by a certain STA to itself to seize the channel before sending a data frame. In this case, the other STAs cannot know which will be the destination of the next frame. Therefore, they should not use the duration field of a CTS for dozing.

\subsection{Practical issues}\label{sec:issues}

\subsubsection{Impact of capture effect}\label{sec:capture-effect}

It is well-known that a high-power transmission can totally blind another one with a lower SNR. Theoretically, two STAs seizing the channel at the same time yields a collision. However, in practice, if the power ratio is sufficiently high, a wireless card is able to decode the high-power frame without error, thus ignoring the other transmission. This is called \emph{capture effect}, and although not described by the standard, it must be taken into account as it is present in real deployments.

According to \cite{Lee2007}, there are two types of capture effect depending on the order of the frames: if the high-power frame comes first, it is called \emph{first} capture effect; otherwise, it is called \emph{second} capture effect. The first one is equivalent to receiving a frame and some noise after it, and then it has no impact in our analysis. In the second capture effect, the receiving STA stops decoding the PLCP of the low-power frame and switches to another with higher power. If the latter arrives \emph{before} a power-saving mechanism makes the decision to go to sleep, the mechanism introduces no misbehaviour.

However, \cite{Lee2007} suggests that a high-power transmission could blind a low-power one \emph{at any time}, even when the actual data transmission has begun. This is called \emph{Message in Message} (MIM) in the literature \cite{mim1, mim2}, and it could negatively impact the performance of an interface implementing an energy-efficiency mechanism based on packet overhearing. In the following, we will provide new experimental evidence supporting that this issue still holds in modern wireless cards.

We evaluated the properties of the MIM effect with an experimental setup consisting of a card under test, a brand new 802.11ac three-stream Qualcomm Atheros QCA988x card, and three additional helper nodes. These are equipped with Broadcom KBFG4318 802.11g cards, whose behaviour can be changed with the open-source firmware OpenFWWF~\cite{openfwwfweb}. We disable the carrier sensing and back-off mechanisms so that we can decide the departure time of every transmitted frame with 1 $\mu$s granularity with respect to the internal 1MHz clock.

\begin{figure}[t]
    \centering
    \includegraphics[width=.8\linewidth]{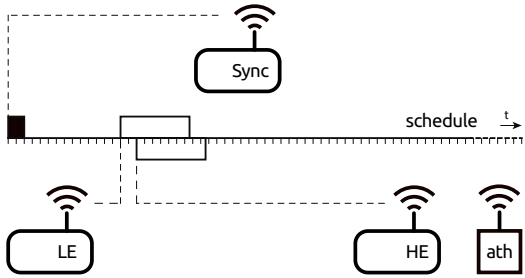}
    \caption{Measurement setup for the MIM effect.}
    \label{fig:secondcapture}
\end{figure}

Fig.~\ref{fig:secondcapture} depicts the measurement setup, which consists of a node equipped with our Atheros card under test (\emph{ath}), a synchronization (Sync) node, a \emph{high energy} (HE) node and a \emph{low energy} (LE) node. These two HE and LE nodes were manually carried around at different distances with respect to the \emph{ath} node until we reached the desired power levels.

The Sync node transmits 80-byte long beacon-like frames periodically at 48 Mbps, one beacon every 8192 $\mu$s: the time among consecutive beacons is divided in 8 schedules of 1024 $\mu$s. Inside each schedule, time is additionally divided into 64 micro-slots of 16  $\mu$s. We then program the firmware of the HE and LE nodes to use the beacon-like frames for keeping their clocks synchronised and to transmit a single frame (138-$\mu$s long) per schedule starting at a specific micro-slot. This allows us to always start the transmission of the \emph{low energy} frame from the LE node before the \emph{high energy} frame from the HE node, and to configure the exact delay $\Delta t$ as a multiple of the micro-slot duration. 

For instance, we set up a $\Delta t = 32$ $\mu$s by configuring LE node to transmit at slot 15, HE node at slot 17. By moving LE node away from the \emph{ath} node while the HE node is always close, we are able to control the relative power difference $\Delta P$ received by the \emph{ath} node between frames coming from the LE and HE nodes. With the configured timings, we are able to replicate the reception experiment at the \emph{ath} node approximately 976 times per second, thus collecting meaningful statistics in seconds. 

We obtained the results shown in Table~\ref{tab:secondcapture}. When the energy gap is small ($\le$ 5 dB), the MIM effect never enters into play as we can see from the first part of Table~\ref{tab:secondcapture}. If the two frames are transmitted at the same time, then the QCA card receives the majority of the HE frames (92\%) despite some of them are broken (17\%); almost no LE frames are received. By increasing the delay to 16 $\mu$s, the QCA card stops working: the short delay means that the HE frame collide with the LE one at the PLCP level. The energy gap does not allow the QCA correlator to restart decoding a new PLCP and, in fact, only a few frames are sporadically received. Further increasing the delay allows the QCA card to correctly receive the PLCP preamble of the LE frame, but then the PDU decoding is affected by errors (e.g., delay set to 48 $\mu$s) because of collision. Finally, if the delay is high enough so that both frames fit into a schedule, the QCA card receives everything correctly ($\ge$ 144 $\mu$s).

When the energy gap exceeds a threshold (e.g., more than 35 dB), then the behaviour of the QCA card changes radically as we can see from the second part of Table~\ref{tab:secondcapture}: first, with no delay, all high energy frames are received (expected given that they overkill the others); second, when both frame types fit in the schedule, all of them are received, which confirms that the link between LE node and the QCA is still very good. But, unlike the previous case, HE frames are received regardless of the delay, which means that the correlator restarts decoding the PLCP of the second frame because of the higher energy, enough for distinguishing it from the first frame that simply turns into a negligible noise.

\begin{table}[t]
	\renewcommand{\arraystretch}{1.3}
	\renewcommand{\tabcolsep}{1.6mm}
	\footnotesize
	\caption{MIM effect.}
	\label{tab:secondcapture}
	\centering
	\begin{tabular}{c|r|rr|rr}
		\hline 
		\multirow{2}{*}{$\Delta P$ [dB]} & \multirow{2}{*}{$\Delta t$ [$\mu$s]} & \multicolumn{2}{c|}{LE frames} & \multicolumn{2}{c}{HE frames}\\
		\cline{3-6} 
 		&& \% rx & \% err & \% rx & \% err\\
		\hline 
		\hline 
		\multirow{5}{*}{$\le$ 5} & 0 & 0.04 & 50.00 & 92.00 & 17.67 \\
		& 16 & 0.40 & 0.00 & 2.15 & 0.00 \\
		& 32 & 99.32 & 99.96 & 0.24 & 0.00 \\
		& $\ge$ 48 & 99.10 & 99.75 & 0.34 & 0.00 \\
		& $\ge$ 144 & 98.94 & 0.00 & 97.32 & 0.00 \\
		\hline
		\multirow{7}{*}{$\ge$ 35} & 0 & 0.18 & 0.00 & 99.37 & 0.00 \\
		& 16 & 0.37 & 11.11 & 91.87 & 0.00 \\
		& 32 & 0.39 & 78.95 & 89.89 & 0.00 \\
		& 48 & 1.54 & 68.00 & 95.58 & 0.00 \\
		& 64 & 3.22 & 98.73 & 89.83 & 0.00 \\
		& 128 & 60.35 & 99.96 & 39.24 & 0.00 \\
		& $\ge$ 144 & 95.33 & 0.00 & 99.64 & 0.00 \\
		\hline 
	\end{tabular}
\end{table}

Thus, our experiments confirm that the MIM effect actually affects modern wireless cards, and therefore it should be taken into account in any micro-sleep strategy. Let us consider, for instance, a common infrastructure-based scenario in which certain STA receives low-power frames from a distant network in the same channel. If the AP does not see them, we are facing the hidden node problem. It is clear that none of these frames will be addressed to our STA, but, if it goes to sleep during these transmissions, it may lose potential high-power frames from its BSSID. Therefore, if we perform micro-sleeps under hidden node conditions, in some cases we may lose frames that we would receive otherwise thanks to the capture effect. The same situation may happen within the local BSSID (the low-power frames belong to the same network), but this is far more rare, as such a hidden node will become disconnected sooner or later.

In order to circumvent these issues, a STA should only exploit micro-sleep opportunities arising from its own network. To discard packets originating from other networks, the algorithm looks at the BSSID in the receiver address within frames addressed to an AP. If the frame was sent by an AP, it only needs to read 6 additional bytes (in the worst case), which are included in the transmitter address. Even so, these additional bytes do not necessarily involve consuming more time, depending on the modulation. For instance, for OFDM 11ag rates, this leads to a time increase of 8 $\mu$s at 6 and 9 Mbps, 4 $\mu$s at 12, 18 and 36 Mbps, and no time increase at 24, 48 and 54 Mbps.

\subsubsection{Impact of errors in the MAC header}\label{sec:probabilistic-analysis}

Taking decisions without checking the FCS (placed at the end of the frame) for errors or adding any protection mechanism may lead to performance degradation due to frame loss. This problem was firstly identified by \cite{Balaji2010} and \cite{Prasad2014} which, based on purely qualitative criteria, reached opposite conclusions. The first work advocates for the need for a new CRC to protect the header bits while the latter dismisses this need. This section is devoted to analyse quantitatively the impact of errors.

At a first stage, we need to identify, field by field, which cases are capable of harming the performance of our algorithm due to frame loss. The duration/ID field (2 bytes) and the MAC addresses (6 bytes each) are an integral part of our algorithm. According to its encoding, the duration/ID field will be interpreted as an actual duration \emph{if and only if the bit 15 is equal to 0}. Given that the bit 15 is the most significant one, this condition is equivalent to the value being smaller than 32 768. Therefore, we can distinguish the following cases in terms of the possible errors:

\begin{itemize}
\item \emph{An error changes the bit 15 from 0 to 1}. The field will not be interpreted as a duration and hence we will not go to sleep. We will be missing an opportunity to save energy, but there will be no frame loss and, therefore, the network performance will not be affected.
\item \emph{An error changes the bit 15 from 1 to 0}. The field will be wrongly interpreted as a duration. The resulting \emph{sleep} will be up to 33 ms longer than required, with the potential frame loss associated.
\item \emph{With the bit 15 equal to 0, an error affects the previous bits}. The resulting \emph{sleep} will be shorter or longer that the real one. In the first case, we will be missing an opportunity to save energy; in the second case, there is again a potential frame loss.
\end{itemize} 
 
Regarding the receiver address field, there exist the following potential issues:

\begin{itemize}
\item \emph{A multicast address changes but remains multicast}. The frame will be received and discarded, i.e., the behaviour will be the same as with no error. Hence, it does not affect.
\item \emph{A unicast address changes to multicast}. The frame will be received and discarded after detecting the error. If the unicast frame was addressed to this host, it does not affect. If it was addressed to another host, we will be missing an opportunity to save energy.
\item \emph{A multicast address changes to unicast}. If the unicast frame is addressed to this host, it does not affect. If it is addressed to another host, we will save energy with a frame which would be otherwise received and discarded.
\item \emph{Another host's unicast address changes to your own}. This case is very unlikely. The frame will be received and discarded, so we will be missing an opportunity to save energy.
\item \emph{Your own unicast address changes to another's}. We will save energy with a frame otherwise received and discarded.
\end{itemize}

As for the transmission address field, this is checked as an additional protection against the undesirable effects of the already discussed intra-frame capture effect. If the local BSSID in a packet changes to another BSSID, we will be missing an opportunity to save energy. It is extremely unlikely that an error in this field could lead to frame loss: a frame from a foreign node (belonging to another BSSID and hidden to our AP) should contain an error that matches the local BSSID in the precise moment in which our AP tries to send us a frame (note that this frame might be received because of the MIM effect explained previously).

Henceforth, we draw the following conclusions from the above analysis:

\begin{itemize}
 \item Errors at the MAC addresses \emph{do not produce frame loss}, because under no circumstances they imply frame loss. The only impact is that there will be several new opportunities to save energy and several others will be wasted.
 \item Errors at the duration/ID field, however, \emph{may produce frame loss} due to frame loss in periods of time up to 33 ms. Also several energy-saving opportunities may be missed without yielding any frame loss.
 \item An error burst affecting both the duration/ID field and the receiver address may potentially change the latter in a way that the frame would be received (multicast bit set to 1) and discarded, and thus preventing the frame loss.
\end{itemize}

From the above, we have that the only case that may yield performance degradation in terms of frame loss is when we have errors in the duration/ID field. In the following, we are going to analytically study and quantify the probability of frame loss in this case. For our analysis, we first consider statistically independent single-bit errors. Each bit is considered the outcome of a Bernoulli trial with a success probability equal to the bit error probability $p_{b}$. Thus, the number of bit errors, $X$, in certain field is given by a Binomial distribution $X\sim \operatorname{B}(N, p_b)$, where $N$ is the length of that field. 

With these assumptions, we can compute the probability of having more than one erroneous bit, $\Pr(X \geq 2)$, which is three-four orders of magnitude smaller than $p_b$ with realistic $p_b$ values. Therefore, we assume that we never have more than one bit error in the frame header, so the probability of receiving an erroneous duration value with a single-bit error, $p_{e,b}$, is the following:
\begin{align}
 p_{e,b} \approx 1 - (1 - p_b)^{15}
\end{align}

However, not all the errors imply a duration value greater than the original one, but only those which convert a zero into a one. Let us call $\operatorname{Hw}(i)$ the Hamming weight, i.e., the number of ones in the binary representation of the integer $i$. The probability of an erroneous duration value greater than the original, $p_{eg,b}$, is the following:
\begin{align}
 p_{eg,b}(i) = p_{e,b}\cdot \frac{15 -\operatorname{Hw}(i)}{15}
\end{align}

\noindent which represents a fraction of the probability $p_{e,b}$ and depends on the original duration $i$ (before the error). 

In order to understand the implications of the above analysis into real networks, we have analysed the data set SIGCOMM'08 \cite{umd-sigcomm2008-2009-03-02} and gathered which duration values are the most common. In the light of the results depicted in Table~\ref{tab:duration}, it seems reasonable to approximate $p_{eg,b}/p_{e,b} \approx 1$, because it is very likely that the resulting duration will be greater than the original.  

\begin{table}[t]
	\renewcommand{\arraystretch}{1.3}
	\footnotesize
	\caption{Most frequent duration values.}
	\label{tab:duration}
	\centering
	\begin{tabu} to \linewidth {r|r|r|X[j]}
		\hline 
		Duration & \% & $p_{eg,b}/p_b$ & Cause \\
		\hline 
		\hline 
		44 & 62.17 & 0.88 & SIFS + ACK at 24 Mbps\\
		0 & 25.23 & 1.00 & Broadcast, multicast... packets\\
		60 & 6.54 & 0.73 & SIFS + ACK at 6 Mbps\\
		48 & 5.82 & 0.87 & SIFS + ACK at 12 Mbps\\
		\hline 
	\end{tabu}
\end{table}

Finally, we can approximate $p_b$ by the BER and, based on the above data and considerations, the frame loss probability, $p_{\mathrm{loss}}$, due to an excessive sleep interval using a single-bit error model is the following:
\begin{align}
 p_{\mathrm{loss}} = p_{eg,b} \approx p_{e,b} \approx 1 - (1 - \mathrm{BER})^{15}
\end{align}

The above analysis assumes that errors occur independently. However, it is well known that in reality errors typically occur in bursts. In order to understand the impact of error bursts in our scheme, we analyse a scenario with independent error bursts of length $X$ bits, where $X$ is a random variable. To this end, we use the Neyman-A contagious model \cite{neyman1939new}, which has been successfully applied in telecommunications to describe burst error distributions \cite{s614, becam1985validite, irvin1991monitoring}. This model assumes that both the bursts and the burst length are Poisson-distributed. Although assuming independency between errors in the same burst may not be accurate, it has been shown that the Neyman-A model performs well for short intervals \cite{cornaglia1996letter}, which is our case.

The probability of having $k$ errors in an interval of $N$ bits, given the Neyman-A model, is the following:
\begin{align}
 p_N(k) = \frac{\lambda_b^k}{k!}e^{-\lambda_B}\sum_{i=0}^\infty\frac{i^k}{i!}\lambda_B^i e^{-i\lambda_b}
\end{align}

\noindent where

\begin{description}
\item[$\lambda_b$] is the average number of bits in a burst.
\item[$\lambda_B$] $= Np_b/\lambda_b$ is the average number of bursts.
\end{description}

This formula can be transformed into a recursive one with finite sums \cite{neyman1939new}:
\begin{align}
 p_N(k) &= \frac{\lambda_B\lambda_b e^{-\lambda_b}}{k}\sum_{j=0}^{k-1} \frac{\lambda_b^j}{j!}p_N(k-1-j) \nonumber\\
 p_N(0) &= e^{-\lambda_B\left(1-e^{-\lambda_b}\right)}
\end{align}

Following the same reasoning as for the single-bit case, we can assume one burst at a time which will convert the duration value into a higher one. Then, the frame loss probability is the following:

\begin{equation}
 p_{\mathrm{loss}} = \sum_{k=1}^{15} p_{15}(k)
\end{equation}

\noindent with parameters $\lambda_b$ and $p_b \approx \mathrm{BER}$.

\begin{figure}[t]
	\centering
	\includegraphics[width=\linewidth]{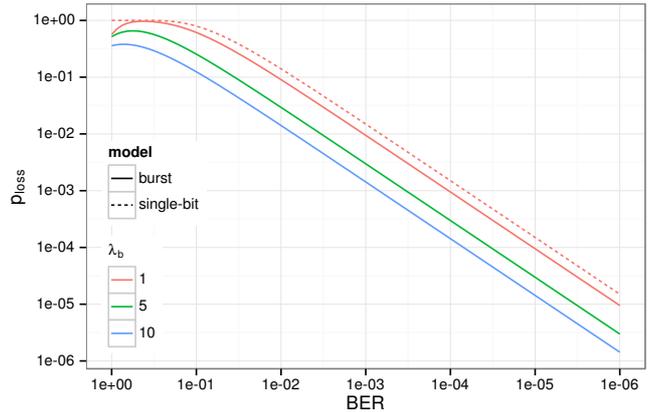}
	\caption{Frame loss probability given a BER level.}
	\label{fig:ploss}
\end{figure}

Fig.~\ref{fig:ploss} evaluates both error models as a function of BER. As expected, the single-bit error model is an upper bound for the error burst model and represents a worst-case scenario. At most, the frame loss probability is one order of magnitude higher than BER. Therefore, we conclude that the frame loss is negligible for reasonable BERs and, consequently, the limited benefit of an additional CRC does not compensate the issues.

\subsection{Algorithm design}\label{sec:design}

In the following, we present \TheName{}, which builds upon the insights provided in previous sections and tries to save energy during the channel transmissions in which the STA is not involved. However, not all transmissions addressed to other stations are eligible for dozing, as the practical issues derived from the capture effect may incur in performance degradation. Therefore, the algorithm must check both the receiver as well as the transmitter address in the MAC header in order to determine whether the incoming frame is addressed to another station \emph{and} it comes from within the same network.

If these conditions are met, a basic micro-sleep will last the duration of the rest of the incoming frame plus an inter-frame space (SIFS). Unfortunately, the long times required to bring an interface back and forth from sleep, as discovered in Section~\ref{sec:transition-times}, shows that this basic micro-sleep may not be long enough to be exploitable. Thus, the algorithm should take advantage of the NAV field whenever possible. Our previous analysis shows that this duration information stored in the NAV is not exploitable in every circumstance: the interface can leverage this additional time during CPs and it must avoid any NAV set by a CTS packet.

\begin{figure}[t]
    \centering
    \includegraphics[width=\linewidth]{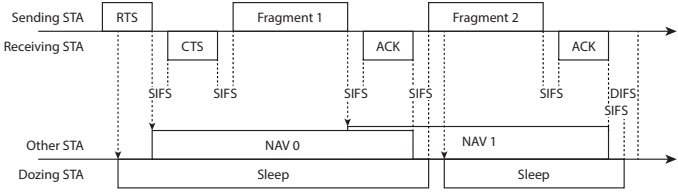}
    \caption{RTS/CTS-based fragmented transmission example and \TheName{}'s behaviour.}
    \label{fig:fragments}
\end{figure}

Finally, after a micro-sleep, two possible situations arise:

\begin{itemize}
 \item The card wakes up at the end of a frame exchange. For instance, after a data + ACK exchange. In this case, all STAs should wait for a DIFS interval before contending again.
 \item The card wakes up in the middle of a frame exchange. For instance, see Fig.~\ref{fig:fragments}, where an RTS/CTS-based fragmented transmission is depicted.
\end{itemize}

In the latter example, an RTS sets the NAV to the end of a fragment, and our algorithm triggers the sleep. This first fragment sets the NAV to the end of the second fragment, but it is not seen by the dozing STA. When the latter wakes up, it sees a SIFS period of silence and then the second fragment, which sets its NAV again and may trigger another sleep. This implies that the STA can doze for an additional SIFS, as Fig.~\ref{fig:fragments} shows, and wait in idle state until a DIFS is completed before trying to contend again.

Based on the above, Algorithm~\ref{alg:unap} describes the main loop of a wireless card's microcontroller that would implement our mechanism. When the first 16 bytes of the incoming frame are received, all the information needed to take the decision is available: the duration value ($\Delta t_\mathrm{NAV}$), the receiver address ($R_A$) and the transmitter address ($T_A$). The ability to stop a frame's reception at any point has been demonstrated to be feasible \cite{berger2014}. Note that MAC addresses can be efficiently compared in a streamed way, so that the first differing byte (if the first byte of the $R_A$ has the multicast bit set to zero, i.e., $R_A$ is unicast) triggers our sleep procedure (\textsc{Set\_Sleep} in Algorithm~\ref{alg:unap}). In addition, the main loop should keep up to date a global variable ($C$) indicating whether the contention is currently allowed (CP) or not (CFP). This is straightforward, as every CFP starts and finishes with a beacon frame.

The \textsc{Set\_Sleep} procedure takes as input the remaining time until the end of the incoming frame ($\Delta t_\mathrm{DATA}$) and the duration value ($\Delta t_\mathrm{NAV}$). The latter is used only if it is a valid duration value and a CP is active. Then, the card may doze during $\Delta t_\mathrm{sleep}$ (if this period is greater than $\Delta t_\mathrm{sleep,min}$), wait for a DIFS to complete and return to the main loop.

Finally, it is worth noting that this algorithm is deterministic, as it is based on a set of conditions to trigger the sleep procedure. It works locally with the information already available in the protocol headers, without incurring in any additional control overhead and without impacting the normal operation of 802.11. Specifically, our analytical study of the impact of errors in the first 16 bytes of the MAC header shows that the probability of performance degradation is comparable to the BER under normal channel conditions. Therefore, the overall performance in terms of throughput and delay is completely equivalent to normal 802.11.

\PSalgorithm{t}

\section{Performance evaluation}\label{sec:results}

This section is devoted to evaluate the performance of \TheName{}. First, Section~\ref{sec:evaluation}, through trace-driven simulation, shows that \TheName{} significantly reduces the overhearing time and the energy consumption of a real network. Secondly, Section~\ref{sec:applicability} analyses the impact of the timing constraints imposed by the hardware, which are specially bad in the case of the AR9280, and discusses the applicability of \TheName{} in terms of those parameters and the evolution trends in the 802.11 standard.

\subsection{Evaluation with real traces}\label{sec:evaluation}

In the following, we conduct an evaluation to assess how much energy might be saved in a real network if all STAs implement \TheName{} using the AR9280, the wireless card characterised in Section~\ref{sec:transition-times}. The reasons for this are twofold. On the one hand, the timing properties of this interface are particularly bad if we think of typical frame durations in 802.11, which means that many micro-sleep opportunities will be lost due to hardware constraints. On the other hand, it does not support newer standards that could potentially lead to longer micro-sleep opportunities through mechanisms such as frame aggregation. Therefore, an evaluation based on an 11a/g network and the AR9280 chip represents a worst case scenario for our algorithm.

For this purpose, we used 802.11a wireless traces with about 44 million packets, divided in 43 files, from the data set SIGCOMM'08 \cite{umd-sigcomm2008-2009-03-02}. The methodology followed to parse each trace file is as follows. Firstly, we discover all the STAs and APs present. Each STA is mapped into its BSSID and a bit array is developed in order to hold the status at each point in time (online or offline). It is hard to say when a certain STA is offline from a capture, because they almost always disappear without sending a disassociation frame. Thus, we use the default rule in \texttt{hostapd}, the daemon that implements the AP functionality in Linux: a STA is considered online if it transmitted a frame within the last 5 min.

Secondly, we measure the amount of time that each STA spends (without our algorithm) in the following states: transmission, reception, overhearing and idle. We consider that online STAs are always awake; i.e., even if a STA announces that it is going into PS mode, we ignore this announcement. We measure also the amount of time that each STA would spend (with our algorithm) in transmission, reception, overhearing, sleep and idle. Transmission and reception times match the previous case, as expected. As part of idle time, we account separately the wasted time in each micro-sleep as a consequence of hardware limitations (the fixed toll $\Delta t_\mathrm{waste}$). After this processing, there are a lot of duplicate unique identifiers (MAC addresses), i.e., STAs appearing in more than one trace file. Those entries are summarised by aggregating the time within each state.

At this point, let us define the \emph{activity} time as the sum of transmission, reception, overhearing, sleep and wasted time. We do not account for idle time since our goal is to understand how much power we can save in the periods of activity, which are the only ones that consume power in wireless transmissions (the scope of this paper). Using the definition above, we found that the majority of STAs reveals very little activity (they are connected for a few seconds and disappear). Therefore, we took the upper decile in terms of activity, thus obtaining the 42 more active STAs.

\begin{figure}[t]
    \centering
    \includegraphics[width=\linewidth]{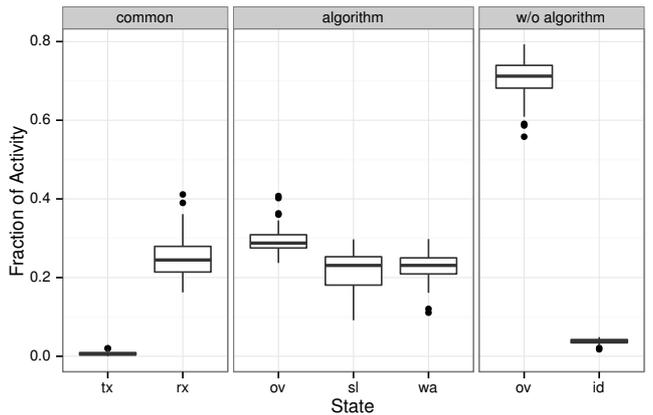}
    \caption{Normalised activity aggregation of all STAs.}
    \label{fig:eval-boxplot}
\end{figure}
\begin{figure}[t]
    \centering
    \includegraphics[width=\linewidth]{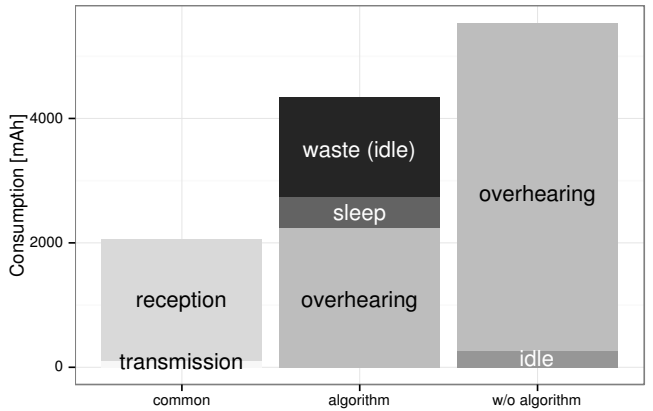}
    \caption{Energy consumption aggregation of all STAs.}
    \label{fig:eval-energy}
\end{figure}

The activity aggregation of all STAs is normalised and represented in Fig.~\ref{fig:eval-boxplot}. Transmission (tx) and reception (rx) times are labelled as \emph{common}, because STAs spend the same time transmitting and receiving both with and without our algorithm. It is clear that our mechanism effectively reduces the total overhearing (ov) time from a median of 70\% to a 30\% approximately (a 57\% reduction). The card spends consistently less time in overhearing because this overhearing time difference, along with some idle (id) time from inter-frame spaces, turns into micro-sleeps, that is, sleep (sl) and wasted (wa) time.

This activity aggregation enables us to calculate the total energy consumption using the power values from the thorough characterisation presented in \ref{sec:characterisation1}. Fig.~\ref{fig:eval-energy} depicts the energy consumption in units of mAh (assuming a typical 3.7-V battery). The energy savings overcome 1200 mAh even with the timing limitations of the AR9280 card, which (1) prevents the card from going to sleep when the overhearing time is not sufficiently long, and (2) wastes a long fixed time in idle during each successful micro-sleep. This reduction amounts to a 21.4\% of the energy spent in overhearing and a 15.8\% of the total energy during the activity time, when the transmission and reception contributions are also considered.

\begin{figure}[t]
    \centering
    \includegraphics[width=\linewidth]{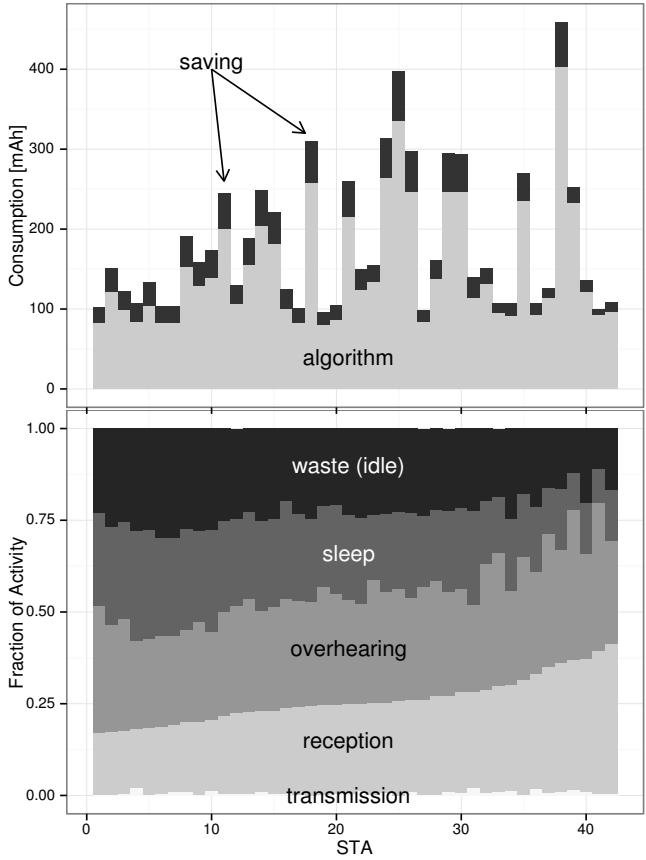}
    \caption{Energy consumption (up) and normalised activity (bottom) for each STA.}
    \label{fig:eval-sta}
\end{figure}

Fig.~\ref{fig:eval-sta} provides a breakdown of the data by STA. The lower graph shows the activity breakdown per STA for our algorithm (transmission bars, in white, are very small). Overhearing time is reduced to a more or less constant fraction for all STAs (i.e., with the algorithm, the overhearing bars represent more or less a 30\% of the total activity for all STAs), while less participative STAs (left part of the graph) spend more time sleeping. The upper graph shows the energy consumption per STA with our algorithm along with the energy-saving in dark gray, which is in the order of tens of mAh per STA.

\subsection{Impact of timing constraints}\label{sec:applicability}

The performance gains of \TheName{} depend on the behaviour of the circuitry. Its capabilities, in terms of timing, determine the maximum savings that can be achieved. Particularly, each micro-sleep has an efficiency (in comparison to an ideal scheme in which the card stays in sleep state over the entire duration of the micro-sleep) given by
\begin{align}
 \frac{E'_\mathrm{save}}{E_\mathrm{save}} = \frac{E_\mathrm{save} - E_\mathrm{waste}}{E_\mathrm{save}} \approx 1 - \frac{\Delta t_\mathrm{waste}}{\Delta t_\mathrm{sleep}}
\end{align}

\noindent which results from the combination of Equations~(\ref{ec:idealsleep}), (\ref{ec:realsleep}) and (\ref{ec:Ewaste}).

\begin{figure}[t]
    \centering
    \includegraphics[width=\linewidth]{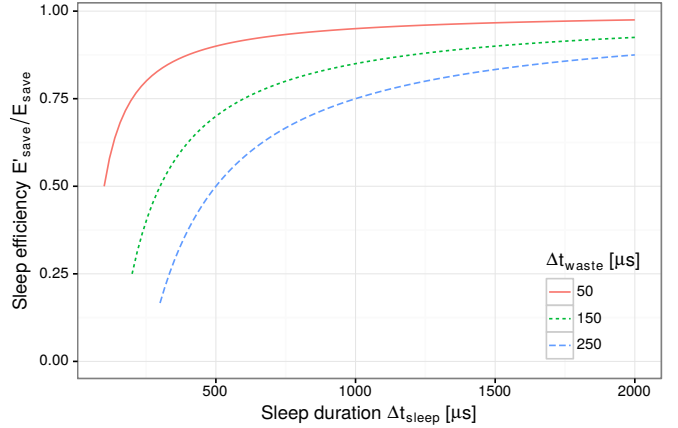}
    \caption{Sleep efficiency behaviour as $\Delta t_\mathrm{waste}$ decreases.}
    \label{fig:savings}
\end{figure}
\begin{figure}[t]
    \centering
    \includegraphics[width=\linewidth]{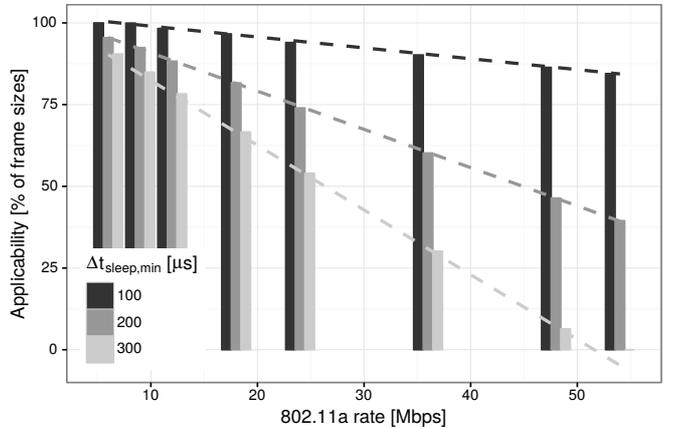}
    \caption{Algorithm applicability for common transmissions ($\le 1500$ bytes $+$ ACK) in 802.11a DCF mode.}
    \label{fig:applicability}
\end{figure}

Fig.~\ref{fig:savings} represents this sleep efficiency for the AR9280 card ($\Delta t_\mathrm{waste}=250$) along with other values. It is clear that an improvement of $\Delta t_\mathrm{waste}$ is fundamental to boost performance in short sleeps.

Similarly, the constraint $\Delta t_\mathrm{sleep,min}$ limits the applicability of \TheName{}, especially in those cases where the NAV cannot be used to extend the micro-sleep. For instance, let us consider the more common case in 11a/b/g networks: the transmission of a frame (up to 1500 bytes long) plus the corresponding ACK. Then,
\begin{align}
 \Delta t_\mathrm{sleep,min} \le \Delta t_\mathrm{DATA} + \Delta t_\mathrm{SIFS} + \Delta t_\mathrm{ACK} + \Delta t_\mathrm{SIFS}
\end{align}

\noindent and expanding the right side of the inequality,
\begin{align}
 \Delta t_\mathrm{sleep,min} &\le \frac{8(14+l_\mathrm{min}+4)}{\lambda_\mathrm{DATA}} + \Delta t_\mathrm{SIFS} \nonumber\\
 &+ \Delta t_\mathrm{PLCP} + \frac{8(14+2)}{\lambda_\mathrm{ACK}} + \Delta t_\mathrm{SIFS}
\end{align}

Here, we can find $l_\mathrm{min}$, which is the minimum amount of data (in bytes, and apart from the MAC header and the FCS) that a frame must contain in order to last $\Delta t_\mathrm{sleep,min}$. Based on this $l_\mathrm{min}$, Fig.~\ref{fig:applicability} defines the applicability in 802.11a DCF in terms of frame sizes ($\le 1500$ bytes) that last $\Delta t_\mathrm{sleep,min}$ at least. Again, an improvement in $\Delta t_\mathrm{waste}$ would boost not only the energy saved per sleep, but also the general applicability defined in this way.

The applicability of \TheName{} may also be affected by the evolution of the standard. Particularly, 802.11n introduced, and 802.11ac followed, a series of changes enabling high and very high throughput respectively, up to Gigabit in the latter case. This improvement is largely based on MIMO and channel binding: multiple spatial and frequency streams. Nevertheless, a single 20-MHz spatial stream is more or less equivalent to 11ag. Some enhancements (shorter guard interval and coding enhancements) may boost the throughput of a single stream from 54 to 72 Mbps under optimum conditions. Yet it is also the case that the PLCP is much longer to accommodate the complexity of the new modulation coding schemes (MCSs). This overhead not only extends each transmission, but also encourages the use of frame aggregation. Thus, the increasing bandwidth, in current amendments or future ones, does not necessarily imply a shorter airtime in practice, and our algorithm is still valid. 

Reducing PHY's timing requirements is essential to boost energy savings, but its feasibility should be further investigated. Nonetheless, there are some clues that suggest that there is plenty of room for improvement. In the first place, $\Delta t_\mathrm{off}$ and $\Delta t_\mathrm{on}$ should depend on the internal firmware implementation (i.e., the complexity of the saved/restored state). Secondly, Fig.~\ref{fig:sleep-tx-a} indicates that a transmission is far more aggressive, in terms of a sudden power rise, than a return from sleep. From this standpoint, $\Delta t_\mathrm{ready} = 200$ $\mu$s would be a pessimistic estimate of the time required by the circuitry to stabilise. Last, but not least, the 802.3 standard goes beyond 802.11 and, albeit to a limited extent, it defines some timing parameters of the PHYs (e.g., $\Delta t_\mathrm{w\_phy}$ would be equivalent to our $\Delta t_\mathrm{on}+\Delta t_\mathrm{ready}$). These timing parameters are in the range of tens of $\mu$s in the worst case (see Table~78-4 of the IEEE Std 802.3-2008 \cite{8023}).

Due to these reasons, WiFi card manufacturers should push for a better power consumption behaviour, which is necessary to boost performance with the power-saving mechanism presented in this paper. Furthermore, it is necessary for the standardisation committees and the manufacturers to collaborate to agree power consumption behaviour guidelines for the hardware (similarly to what has been done with 802.3). Indeed, strict timing parameters would allow researchers and developers to design more advanced power-saving schemes.

\section{Conclusions}\label{sec:conclusions}

Based on a thorough characterisation of the timing constraints and energy consumption of 802.11 interfaces, we have exhaustively analysed the micro-sleep opportunities that are available in current WLANs. We have unveiled the practical challenges of these opportunities, previously unnoticed in the literature, and, building on this knowledge, we have proposed \TheName{} an energy-saving scheme that is orthogonal to the existing standard PS mechanisms. Unlike previous attempts, our scheme takes into account the non-zero time and energy required to move back and forth between the active and sleep states, and decides when to put the interface to sleep in order to make the most of these opportunities while avoiding frame losses.

We have demonstrated the feasibility of our approach using a robust methodology and high-precision instrumentation, showing that, despite the limitations of COTS hardware, the use of our scheme would result in a 57\% reduction in the time spent in overhearing, thus leading to an energy saving of 15.8\% of the activity time according to our trace-based simulation. Finally, based on these results, we have made the case for the strict specification of energy-related parameters of 802.11 hardware, which would enable the design of platform-agnostic energy-saving strategies.

\appendix

\section{Energy consumption characterisation}\label{sec:characterisation}

\subsection{State consumption parametrisation}\label{sec:characterisation1}

In order to gain insight into the energy savings of \TheName{}, we performed a complete state parametrisation (power consumption in transmission, reception, overhearing, idle and sleep) of the AR9280 card (the active state in the traces used for the evaluation, Section~\ref{sec:evaluation}) using the same scenario as in Section~\ref{sec:transition-times} (see Fig.~\ref{fig:setup}). As in Section~\ref{sec:transition-times}, all measurements (except for the sleep state) were taken with the wireless card associated to the AP in 11a mode to avoid any interfering traffic, and it was placed very close to the node to obtain the best possible signal quality. The reception of beacons is accounted in the baseline consumption (idle). 

The card under test performed transmissions/receptions to/from the AP at a constant rate and with fixed packet length. In order to avoid artifacts from the reception/transission of ACKs, UDP was used and the NoACK policy was enabled. Packet overhearing was tested by generating traffic of the same characteristics from a secondary STA placed in the same close range ($\sim$cm). Under these conditions, several values of airtime percentage were swept. For each experiment, current and voltage signals were sampled at 100 kHz and the mean power consumption was measured with a basic precision of 1 mW over intervals of 3 s.

Regarding the sleep state, the driver \texttt{ath9k} internally defines three states of operation: \emph{awake}, \emph{network sleep} and \emph{full sleep}. A closer analysis reveals that the card is \emph{awake}, or in \emph{active state}, when it is operational (i.e., transmitting, receiving or in idle state, whether as part of an SSID or in monitor mode), and it is in \emph{full sleep} state when it is not operational at all (i.e., interface down or up but not connected to any SSID). The \emph{network sleep} state is used by the 802.11 PS mechanism, but essentially works in the same way as \emph{full sleep}, that is, it turns off the main reference clock and switches to a secondary 32 kHz one. Therefore, we saw that \emph{full sleep} and \emph{network sleep} are the same state in terms of energy: they consume exactly the same power. The only difference is that \emph{network sleep} sets up a tasklet to wake the interface periodically (to receive TIMs), as required by the PS mode.

\begin{figure}[t]
    \centering
    \includegraphics[width=\linewidth]{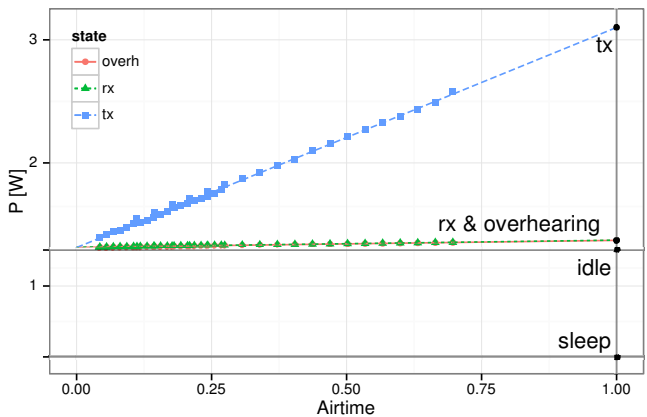}
    \caption{Atheros AR9280 power consumption in 11a mode.}
    \label{fig:power}
\end{figure}

Fig.~\ref{fig:power} shows our results for transmission, reception and overhearing. Idle and sleep consumptions were measured independently, are depicted with gray horizontal lines for reference. As expected, power consumptions in transmission/reception/overhearing state are proportional to airtime, thus the power consumption of such operations can be easily estimated by extrapolating the regression line to the 100\% of airtime (gray vertical line).

These mean values are shown in Table~\ref{tab:power}. First of all, reception and overhearing consumptions are the same within the error, and they are close to idle consumption. Transmission power is more than two times larger than reception. Finally, the sleep state saves almost the 70\% of the energy compared to idle/reception.

\begin{table}[t]
	\renewcommand{\arraystretch}{1.3}
	\footnotesize
	\caption{Atheros AR9280 power consumption}
	\label{tab:power}
	\centering
	\begin{tabular}{r|c|cc|l}
		\hline 
		State & Mode & Channel & MHz & Power [W] \\
		\hline 
		\hline 
		Transmission & 11a & 44 & 20 & 3.10(2)\\
		Reception & 11a & 44 & 20  & 1.373(1)\\
		Overhearing & 11a & 44 & 20  & 1.371(1)\\
		Idle & 11a & 44 & 20 & 1.292(2)\\
		Sleep & - & - & - & 0.424(2)\\
		\hline
		Idle & 11n & 11 & 20  & 1.137(4)\\
		Idle & 11n & 11 & 40  & 1.360(4)\\
		\hline 
	\end{tabular}
\end{table}

\subsection{Downclocking consumption characterisation}\label{sec:characterisation2}

As the AR9280's documentation states, its reference clock runs at 44 MHz for 20 MHz channels and at 88 MHz for 40 MHz channels in the 2.4 GHz band, and at 40 MHz for 20 MHz channels and at 80 MHz for 40 MHz channels in the 5 GHz band. Thus, as Table~\ref{tab:power} shows, we measured two more results to gain additional insight into the behaviour of the main reference clock, which is known to be linear \cite{Zhang2012}.

Using an 11n-capable AP, we measured the idle power in the 2.4 GHz band with two channel widths, 20 and 40 MHz. Note that the idle power in 11a mode (5 GHz band), with a 40 MHz clock, is higher than the idle power with a 44 MHz clock. This is because both bands are not directly comparable, as the 5 GHz one requires more amplification (the effect of the RF amplifier is out of the scope of this paper).

With these two points, we can assume a higher error (of about 10 mW) and try to estimate a maximum and a minimum slope for the power consumed by the main clock as a function of the frequency $f$. The resulting averaged regression formula is the following:
\begin{align}
 P(f) = 0.91(3) + 0.0051(5)f
\end{align}

This result, although coarse, enables us to estimate how a downclocking approach should perform in COTS devices. It shows that the main consumption of the clock goes to the baseline power (the power needed to simply turn it on), and that the increment per MHz is low: 5.1(5) mW/MHz. As a consequence, power-saving mechanisms based on idle downclocking, such as \cite{Zhang2012}, will not save too much energy compared to the sleep state of COTS devices. For instance, the x16 downclock of \cite{Zhang2012} applied to this Atheros card throws an idle power consumption of 1.10(2) W in 11a mode, i.e., about a 15\% of saving according to Table~\ref{tab:power}, which is low compared to the 70\% of its sleep state. This questions the effectiveness of complex schemes based on downclocking compared to simpler ones based on the already existing sleep state.

\newpage
\bibliographystyle{elsarticle-num}
\bibliography{unap}

\end{document}